\documentclass[journal=jacsat,manuscript=article]{achemso}

\usepackage[version=3]{mhchem} % Formula subscripts using \ce{}
\usepackage[T1]{fontenc}       % Use modern font encodings
\usepackage{amsmath}
\usepackage{soul,xcolor}
\author{Sebastian K.H. Andersen}
\email{sekh@mci.sdu.dk}
\author{Shailesh Kumar}
\author{Sergey I. Bozhevolnyi}
\affiliation[Unknown University]
{Center for Nano Optics, University of Southern Denmark, Campusvej 55, DK-5230 Odense M, Denmark}
\title[An \textsf{achemso} demo]
  {Ultrabright Linearly Polarized Photon Generation from a Nitrogen Vacancy Center in a Nanocube Dimer Antenna}

\abbreviations{NV-center, Nitrogen vacancy center; ND, nano diamond; QE, quantum emitter; AFM, atomic force microscope; APD, avalance photo diode}

\begin{document}

\begin{tocentry}

%\begin{figure}[H]
%	\centering
		%\includegraphics[width=0.60\textwidth]{Figures/LocalCoordinates.png}
		\begin{center}
		\includegraphics{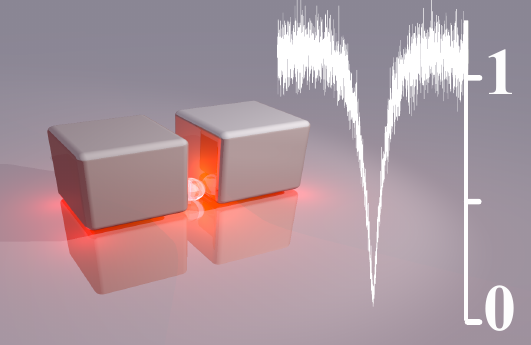}
		\end{center}
%	\label{fig:Figure4}
%\end{figure}

\end{tocentry}

\begin{abstract}
We demonstrate an exceptionally bright photon source based on a single nitrogen-vacancy center (NV-center) in a nanodiamond (ND), placed in the nanoscale gap between two monocrystalline silver cubes in a dimer configuration. The system is operated near saturation at a stable photon rate of 850\,kcps, while we further achieve strongly polarized emission and high single photon purity, evident by the measured auto-correlation with a g$^{(2)}$(0)-value of 0.08. These photon source features are key parameters for quantum technological applications, such as secure communication based on quantum key distribution. The cube antenna is assembled with an atomic force microscope, which allows us to predetermine the dipole orientation of the NV-center and optimize cube positioning accordingly, while also tracking the evolution of emission parameters from isolated ND to the 1 and 2 cube configuration. The experiment is well described by finite element modelling, assuming an instrinsic quantum efficiency of 0.35. We attribute the large photon rate of the assembled photon source, to increased quantum efficiency of the NV-center and high antenna efficiency.
\end{abstract}

\section{Introduction}
The single photon source is a fundamental component in the quantum technological evolution\cite{QuantTechReview}, enabling secure communication based on single photon quantum key distribution \cite{QuantumKey,NVCryptography}, optical quantum computing\cite{LinearQuantumComputing} or light-matter interfaces for quantum infomation storage\cite{QuantumRegister}. For practical applications the single photon source should efficiently deliver a stable stream of indistinguisble, strongly polarized single photons at high rate and high purity, preferably under room temperature operation. Single photon emission may be achieved by spontaneous emission from a wealth of quantum emitters (QE) such as a molecule, quantum dot, defect in diamond, boron nitride or carbon nanotubes etc. \cite{SolidStateEmitterRev}. However the particular single photon emission properties are a function of both the intrinsic characteristics of the emitter and the surrounding dielectric enviroment. The desirable single photon source properties are hence achieved by proper choice of QE and engineering of the photonic enviroment, i.e. by coupling the QE to an antenna\cite{SpecModulation,PolarizationModulation,EmissionPattern}, waveguide\cite{VGroove-NV,NDonSilverWire} or directing the emission with a solid immersion lens. For photostable operation at room-temperature the nitrogen-vacancy center in diamond (NV-center) is an excellent emitter\cite{NVPhotoStable}, while the emission is poorly polarized given phonon promoted population averaging of the doublet excited state\cite{ExcitedStateTimeAvg1,ExcitedStateTimeAvg2}. Several works have previously focused on improving the efficiency and photon rate of the NV-center either by top-down fabrication in bulk diamond\cite{DiamondPillar,DiamondAgAperture} or by incorporating a nanodiamond (ND) containing an NV-center into a plasmonic antenna\cite{GoldSphereDimer}. Though often, engineering the photonic enviroment comes at the cost of reducing the purity of the source, as background fluorescence from impurities in the diamond or materials introduced during fabrication limit the single photon quantum character of the source. The degree to which background compromises the single photon character, may be measured by the dip of the auto-correlation function at time zero (g$^{(2)}(0)$), indicating the probability of detecting 2 photons at the same time. The measured g$^{(2)}(0)$-value of previously demonstrated  NV-center based photon sources have typically been $\sim$ 0.3\cite{DiamondAgAperture,DiamondMembraneBullseye,AgGratingAperture,GoldSphereDimer}       
Schietinger \textit{et. al.} assembled a dimer antenna consisting of two gold spheres around a ND containing an NV-center and demonstrated photon rates up to $\sim$420\,kcps with g$^{(2)}(0)=$0.3\cite{GoldSphereDimer}. Choy \textit{et. al.} realized an NV-center in a diamond post surrounded by a silver film with grating corrugations and reported an asymptotic photon rate limit up to 704\,kcps, achievable at infinite laser power, with g$^{(2)}(0)\sim0.2$\cite{AgGratingAperture}. A record asymptotic limit of 2.7\,Mcps from an NV-center in a diamond membrane with an etched in grating was reported by \textit{Li et. al.} with g$^{(2)}(0)=0.28$\cite{DiamondMembraneBullseye}\\\\

In this work, we demonstrate an exeptionally pure photon source, consisting of a nanodiamond containing a single NV-center situated in the gap between two monocrystalline silver nanocubes. We measure a g$^{2}(0)$-value of 0.08 for the total system. The configuration is stable under large pump powers, as we operate the source near saturation, at a detected photon rate of 850\,kcps, similar to state of the art NV-center sources based on solid immersion lenses in diamond ($\sim$ 1\,Mcps)\cite{DiamondSolidImmersionLens} or ZrO$_2$ ($\sim$ 850\,kcps, stable emission)\cite{ZrOSolidImmersionLens}. Though unlike such sources, we further demonstrate strongly linearly polarized emission from our NV-center with a polarization ratio of 9, between the power detected along the major and minor axis. The system is assembled with an atomic force microscope (AFM), which allows us to directly probe the evolution of the NV-center emission properties from the isolated ND to the 1 cube- and 2 cube configuration. The freedom to precharacterize our NV-center before system assembly allows us to determine the orientation of the NV-center dipole axes and optimize the position of our nanocubes accordingly, for optimal NV-to-antenna coupling. Such optimization improves both the excitation efficiency and photon rate of the NV-center, as the excitation rate is improved by a factor of 5.86, while the detected photon rate at saturation increases by a factor 6.6, relative to the isolated ND. Finite element modelling agree well with experimental observations assuming an intrinsic quantum efficiency of 0.35. The enhanced photon rate, detected from the assembled photon source, is attributed to increased quantum efficiency and high radiation efficiency of the antenna.

\section{Results and Discussion}
\begin{figure}[H]
	\centering
		\includegraphics{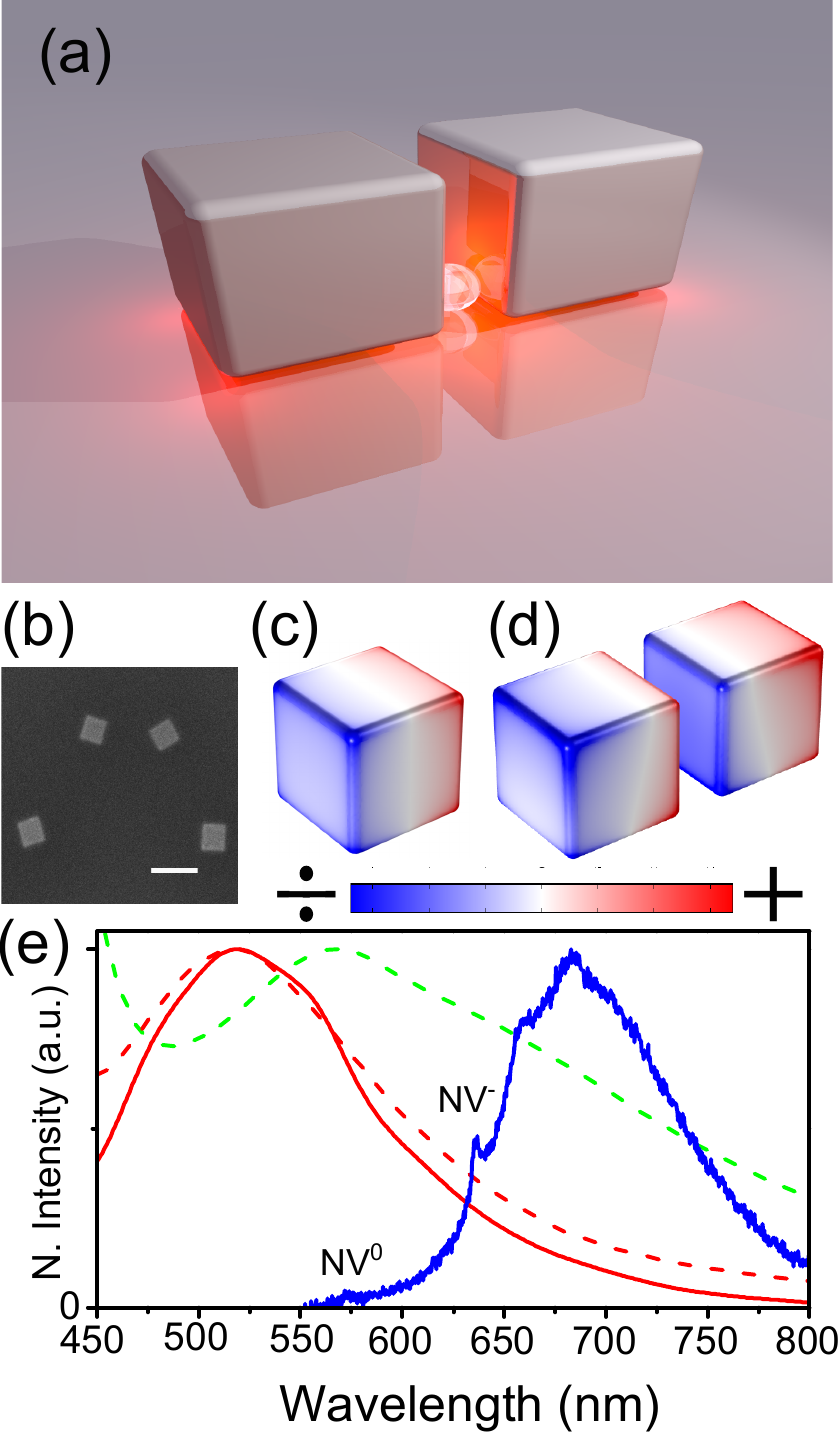}
		\caption{(a) Illustration of ND containing a single NV-center situated in the gap between two silver nano cubes. The NV-center emission is accelerated as it couples to the plasmonic cube mode.(b) Electron micrograph of silver cubes scalebar 200\,nm. (c, d) Charge distribution of the dipolar mode of a single and coupled cubes.(e) Fluorescence spectrum of single NV-center(blue) and measured scattering spectrum of single nanocube $\sim$\,100nm(red) with corresponding simulations for single (red dashed) and coupled cubes(green dashed) separated by a 35\,nm gap.}
	\label{fig:Figure1}
\end{figure}

Figure \ref{fig:Figure1}a illustrates the physics of our experiment. A single NV-center in a ND, decays from the excited electronic state by excitation of plasmonic charge density oscillations in a single- (figure \ref{fig:Figure1}c) or 2 cube configuration (figure \ref{fig:Figure1}d). The excited antenna mode subsequently decays either by ohmic loss or scattering of single photons, polarized along the axis of charge oscillation, hence imposing the radiation properties of the antenna on the NV-center. Further, the photon rate of the system is increased, as the NV-center spends less time in the excited state between emission events, when allowed to efficently dissipate its energy into the antenna mode. Clearly, these desirable single photon source features rely on achieving a large NV-to-antenna coupling rate, compared to direct photon emission or metal quenching. Optimizing the NV-to-antenna coupling rate is typically done by maximizing the quality factor to mode volume ratio of the antenna, spectrally tuned for the optical transition. The NV-center should further be positioned at the point of maximum mode amplitude, with the NV-dipole axis coaligned with the eletric field of the antenna mode \cite{OptimizingCouplingRate}. The near-field interaction of two cubes hybridize the dipolar cube modes (figure \ref{fig:Figure1}c) into an "antibonding" and a "bonding" mode (figure \ref{fig:Figure1}d) shifted to respectively higher and lower energies (figure \ref{fig:Figure1}e). The "bonding" mode is particular well-suited for coupling to a QE, as the mode is strongly confined to the nano scale gap\cite{DimerFieldEnhancement}, while the superradiant damping of the in-phase cube oscillations, ensure high antenna radiation efficiency\cite{DimerEfficiency}. The capacitor-like field distribution in the gap requires the cube facet to be aligned normal to the dipole axes of the NV-center for optimal coupling. We realize the optimal coupling configuration by precharacterization of the NV-center dipole orientation and deterministic nanoassembly of the cube dimer antenna with an AFM. \\
The NV-center is pumped with a 532\,nm linearly polarized continous wave or pulsed laser. The pump light is focused onto the ND by an oil immersion objective (NA 1.4), situated below the quartz glass sample. Photons spontanously emitted into the same objective are filtered by a dichroic mirror (cut-off 550\,nm) and detected by two avalanche photo diodes (APD) in a Hanbury Brown-Twiss configuration, a grating spectrometer or a CCD camera. The pump polarization is controlled by a halfwave plate in the excitation light path, while an analyzer introduced in the detection path probe the polarization of emitted photons. Nano particle manipulation is performed by an AFM positioned above the sample (see supporting information (SI) for a schematic). Pumping a $\sim$ 35\,nm ND (figure \ref{fig:Figure2}b), we observe single photon emission from an NV-center, identified by a measured g$^{(2)}(0)$-value of 0.14 (Figure \ref{fig:Figure3}c) and the zero phonon line fingerprint for the neutral(NV$^{0}$, 575\,nm) and negative charge state(NV$^{-}$, 637\,nm)(figure \ref{fig:Figure1}e). The NV-center continuously flip-flop between the charge states by photoionization\cite{PhotoIonization}, however photon emission is dominated by the NV$^{-}$-state, populated $\sim$ 75\% of the time. Our experiment is hence well described by considering purely the NV$^{-}$-state (see SI for cube coupled NV-center spectra). Excitation and emission from the excited doublet state of NV$^{-}$-center is facilitated by two orthogonal dipole axes (\textbf{p}$_x$, \textbf{p}$_y$) lying in the plane normal to the nitrogen atom - vacancy axis \cite{DipoleOrientation} (figure \ref{fig:Figure2}a). We probe the projection of \textbf{p}$_x$, \textbf{p}$_y$ on the sample plane by rotating the pump polarization, while operating the NV-center in the non-saturated regime (105\,$\mu$W pump power) (figure \ref{fig:Figure2}e). The detected photon rate then follow $R(\varphi)=\eta{q_e}\gamma_{ex}(\varphi)$. $\eta$ being the collection effiency of the objective and q$_e$ the quantum efficiency of the emitter. The excitation rate $\gamma_{ex}(\varphi) \propto \sum_{i=x,y} |\textbf{p}_i\cdot{\textbf{E}}(\varphi)|^2$ depends on the electric field of the pump laser $\textbf {E}=\textbf{E}_{pump}$, at the emitter position, polarized along the sample plane at the azimuth angle $\varphi$. We fit $R(\varphi)$ for the dipole plane containing $\textbf{p}_x$, $\textbf{p}_y$ and determine the largest dipole projection on the sample plane by the pump angle $\varphi^{NV}_p$ resulting in maximum photon rate (figure \ref{fig:Figure2}e). $\varphi^{NV}_p$ is indicated by the blue arrow, wrt. the system configurations in figure \ref{fig:Figure2}b-d. Two 80\,nm chemically synthesized silver cubes are subsequently positioned along $\varphi^{NV}_p$ in a dimer configuration, for optimal NV-to-antenna coupling (figure \ref{fig:Figure2}c, d). $R(\varphi)$ for the cube-coupled system is well described by adding an additional term for the electric field generated by the cube(s) $\textbf{E}_{cube}=f_e\overset{\text{\scriptsize$\leftrightarrow$}}{\alpha}\textbf{E}_{pump}$, such that $\textbf{E}=\textbf{E}_{pump}+\textbf{E}_{cube}$. $f_e$+1 being the electric field enhancement and $\overset{\text{\scriptsize$\leftrightarrow$}}{\alpha}(\varphi^{cube}_p)$ the polarizability tensor of the cube, defined by the orientation of the cube dipole moment ($\varphi^{cube}_p$), which is coaligned with the electric field of the cube mode. We account for the enhancement factor $\eta{q}_e/\eta_0{q}_{e,0}$ of 1.52 (1 cube) and 2 (2 cubes), relative to the isolated ND indexed 0, and fit f$_e$, $\varphi^{cube}_p$ to experiment (figure \ref{fig:Figure2}f, g)(see SI for details). For both cube configurations we find $\varphi^{cube}_p$-$\varphi^{NV}_p$=15$^o$, confirming near optimal alignment of the NV dipole axes with the electric field of the cube(s). The orientation of $\varphi^{cube}_p$ wrt. cube configurations is given by respectively a red or green arrow in figure \ref{fig:Figure2}c, d. The enhancement of excitation rate is determined by $\gamma_{ex}/\gamma_{ex,0}=Rq_{e0}\eta_0/R_{0}q_{e}\eta$ for which we find values of 5.1 (1 cube) and 5.86 (2 cubes) at pump orientation $\varphi^{cube}_p$. Having established the dipole orientation of the NV-center and optimized the antenna configuration accordingly, we turn to the emission properties of the photon source. Rotating an analyzer in front of our detector, we find weakly polarized photons emitted from the isolated ND, with a polarization ratio of r$_{pol}$=2.1, between the photon rate detected along the major and minor axis (figure \ref{fig:Figure2}h). The major axis of polarization is coaligned with $\varphi^{NV}_p$, as photons are generally polarized along the dipole axis of emission, while the two dipole configuration of the NV$^{-}$-state result in an overall weak photon polarization. The photon emission becomes increasingly polarized throughout assembly of the cube antenna, as we find r$_{pol}=$6.9 for a single cube and r $_{pol}$=9 for two cubes (figure \ref{fig:Figure2}i, j). We note a slight shift of the major axis of polarization toward $\varphi^{cube}_p$, in good agreement with the polarization of emission, being the result of polarized photon scattering from plasmonic charge oscillations along the antenna dipole axis. The increase of photon polarization is naturally accompanied by an increase of the excited state decay rate ($\gamma$), as the degree of photon polarization scales with the increasing rate at which the NV-center decays from the excited state, by driving charge oscillations in the cube antenna. We recorded the temporal response to a sharp excitation pulse and found an enhancement of the excited state decay rate of respectively $\gamma{/}\gamma_0=2.25$ for 1 cube and $\gamma{/}\gamma_0=3.28$ for 2 cubes (figure \ref{fig:Figure3}a). Unfortunately, the faster decay rate does not translate directly to an increase in photon rate, as the decay rate enhancement may partially result from metal quenching, or emission coupled to the antenna may be lost ohmically. The brightness of the source is hence determined by tracing out the saturation curve in terms of detected photon rate as a function of pump power (figure \ref{fig:Figure3}b). After substraction of background from the plain sample surface, the detected photon rate is fitted to the conventional model:

\begin{align}
R(P)=R_{\infty}\frac{P}{P+P_{sat}}
\end{align}

P being the pump power, P$_{sat}$ the saturation power and $R_{\infty}$ the asymptotic photon rate limit, detected at infinite pump power. Curiously, the R$_{\infty}$/R$_{\infty{0}}$ enhancement is significantly larger than the decay rate enhancement, as we find R$_{\infty}$/R$_{\infty{0}}$=3.41 for 1 cube and R$_{\infty}$/R$_{\infty{0}}$=6.58 for two cubes, with R$_{\infty}$=914\,kcps. The system is stable under large pump power as we operate the source at a photon rate of 850\,kcps, close to the photon rate limit. An AFM scan after $\sim$10\,min of operation indicates no morphological changes of the antenna. We attribute such power stability to the low ohmic heating losses of pristine monocrystalline silver, impeding thermal deformation even at large pump powers, while the strong radiative damping of the dimer mode may also be a contributing factor. Further, the large interband transision energy of silver $\sim$3.4\,eV\cite{SilverBand}, prevents background photoluminescence from the metal, thereby ensuring high single photon purity. The photon purity is examined by histogramming the time interval ($\tau$) between photon detection events, yielding the 2. order correlation function g$^{(2)}(\tau)$. g$^{(2)}(0)$-events hence correspond to simultaneous detection of 2 photons, only possible in the presence of background as the NV-center may only emit 1 photon at a time. The g$^{(2)}(0)$-value is determined by fitting a 3-level rate model including a background term, normalized for $\tau\rightarrow\infty$\cite{HechtNovotny} (figure \ref{fig:Figure3}c-e). The g$^{(2)}(0)$-value slightly improves by the addition of cubes to 0.08 from 0.14 measured for the isolated ND. The improved g$^{(2)}(0)$-value may be a result of the increased brightness of the NV-center effectively improving the signal-to-background ratio.\\ 
Concluding the description of photon source emission properties, we now numerically examine the relation of the experimentally observed enhancement factors $\gamma/\gamma_0$ and R$_{\infty}$/R$_{\infty{0}}$. We write the decay rate of the isolated NV-center in terms of a radiative rate ($\gamma_{r0}$) and an intrinsic non-radiative rate ($\gamma_{nr0}$), such that $\gamma_0=\gamma_{r0}+\gamma_{nr0}$ with a corresponding intrinsic quantum efficiency $q_{e0}=\gamma_{r0}/\gamma_{0}$. Introducing a silver cube accelerates the radiative decay rate as additional decay channels, such as the plasmonic cube mode or metal quenching, are available for emission, while $\gamma_{nr0}$ is unaffected by the enviroment. The changed radiative decay is written $\gamma_r=\Gamma\gamma_{r0}$ such that the decay rate for the one or two cube configuration is given by $\gamma=\Gamma\gamma_{r0}+\gamma_{nr0}$. The decay rate enhancement in this case takes the form:

\begin{align}
\frac{\gamma}{\gamma_0}=\Gamma{q_{e0}}+(1-q_{e0})
\label{eq:DecayRateEnh}
\end{align}

The photon rate limit is given by $R_{\infty}=\eta\gamma_r$. We define collection efficiency as the probability of a radiative decay event, resulting in a photon being emitted into the objective. Losses due to metal quenching or ohmic dissipation of antenna excitations are thereby included in $\eta$ and the photon rate enhancement can be written:

\begin{align}
\frac{R_\infty}{R_{\infty{0}}}=\frac{\eta}{\eta_0}\Gamma
\label{eq:PhotonRateEnh}
\end{align}

The NV-center emission is modelled by 3D finite element simulations, as the power dissipated at 680\,nm wavelength by two orthogonal electric dipoles, oriented along the experimentally determined dipole plane and positioned in the center of a 35\,nm ND. The ND is modelled as a 4-sided truncated pyramid.  $\Gamma$ is then obtained as the total power dissipated in the one or two cube configuration, relative to the ND situated on a glass substrate. Decay rate enhancement is calculated by equation \ref{eq:DecayRateEnh} for various q$_{e0}$ values (figure \ref{fig:Figure4}a). We find good agreement of modelled and experimental values for realistic gap sizes 40-45\,nm, for an intrinsic quantum efficiency of q$_{e0}\sim{0.35}$, in good agreement with previous experimental studies of the ND product\cite{NDQuantumEff}. Setting q$_{e0}=0.35$ and inserting experimental values $\gamma/\gamma_0$, $R_\infty/R_{\infty{0}}$ in equation \ref{eq:DecayRateEnh} and \ref{eq:PhotonRateEnh}, we confirm the consistency of experiment and model by predicting $\eta/\eta_0$ in similar good agreement with modelling (figure \ref{fig:Figure4}b (inset)), while $R_{\infty}/R_{\infty{0}}$ may be directly modelled for similar gap sizes (figure \ref{fig:Figure4}b). The disparity of $\gamma/\gamma_0$ and $R_\infty/R_{\infty{0}}$ is thereby well explained by the non-unity instrinsic quantum efficiency of the NV-center. For q$_{e0}$=0.35 we find $\Gamma=4.57$ for one cube and $\Gamma=7.51$ for 2 cubes using equation 2. The corresponding quantum efficiency is calculated to respectively 0.71 and 0.80 by $q_e=\Gamma{q_{e0}}/(1+q_{e0}(\Gamma-1))$. The modelled and experimentally predicted increase of collection efficiency, going from one to two cubes (figure \ref{fig:Figure4}b (inset)), is attributed to an increase of antenna efficiency as the numerical model finds respectively $\sim$ 63\,$\%$ and $\sim$ 97\,$\%$ of dipole emission reaching the far-field, while the fraction  of far field emssion collected by the objective, is nearly unchanged. The increase in antenna efficiency should be expected given the superradiative damping of the dimer mode, compared to the plasmonic mode of a single nano particle. The large photon enhancement is thereby a result of an increased quantum efficiency of the NV-center, while an improved antenna efficiency of the dimer configuration,  is a contributing factor to the enhancement over the single cube.\\
We conducted the experiment 4 times, labelled experiment A-D, A being the experiment presented up to this point(figure \ref{fig:Figure4}c-f) (see SI for measurements). Consistently we find $R_{\infty}/R_{\infty{0}} > \gamma/\gamma_{0}$ suggestive of a non-unity intrinsic quantum efficiency. The photon rate limit improved for all assembled photon sources by the addition of a second cube, with values $R_{\infty}$=671 - 1460\,kcps and a consistently high photon purity of g$^{(2)}(0)$=0.08 - 0.26 for the dimer configuration (figure \ref{fig:Figure4}c, d). Enhancement factors for a single cube are $\gamma/\gamma_0$=1.8 - 3.4 ; $R_{\infty}/R_{\infty{0}}$=2.1 - 4.0, while the spread is more significant for two cubes $\gamma/\gamma_0$=2.1 - 5.9 ; $R_{\infty}/R_{\infty{0}}$=2.7 - 18.0\,. The larger spread is expected, given the strong dependence on cube separation, which is limited to the size of the ND, varying between experiments $\sim$30-35\,nm. Asymmetry of cube configurations and varying dipole orientation, and position of NV-center in the ND, should further contribute to the spread in experimental results. 

\section{Conclusion and outlook}
In summary we have presented a remarkably pure photon source, with a g$^{(2)}(0)$-value of 0.08, based on an NV-center contained in a $\sim$ 35\,nm ND, placed in the gap between two monocrystalline silver nano cubes. The low ohmic heating losses of pristine monocrystalline silver, allowed for stable operation under large laser powers, at a detected photon rate of 850\,kcps near the saturation limit of 914\,kcps. We demonstrated how AFM assembly of the photon source allowed for near optimal alignment of nano cubes for the maximum inplane dipole moment of the NV-center, while futher tracking the photon polarization properties, from weak polarization for the bare ND to strong linear polarization for the assembled system.
The experimental finds is consistent with modelling of an NV-center with an instrinsic quantum efficiency of $\sim$ 0.35. The presented results are quite encouraging as significant improvements is within reach, by going for smaller gap sizes by employing smaller ND's down to 5\,nm in size, for stable NV-center emission\cite{NVND} or 1.6\,nm for the Silicon vacancy center\cite{SiVND}. A scalable approach to realizing the photon source, is also concievable with molecular self-assembly as controlled assembly of face-face or egde-edge cube dimer configurations have already been realized.\cite{CubeAssembly}.

\section{Methods}
\subsection{Sample preparation}
A 0.18\,mm thick fused quartz glass slide (SPI supplies) was cleaned by an RCA1 cleaning step. 5\,ml Mili-Q and 1\,ml 28-30\,$\%$ NH$_4$OH(aq) solution heated to 65\,C, were removed from heating plate and 1\,ml 30\,$\%$ H$_2$O$_2$(aq) and slide glass added for 10\,min, followed by 2 step submersion in Mili-Q baths, 5\,min each. The cleaning step removed organic residue and promoted surface hydrophilicity for subsequent spincoating of a ND solution <50\,nm (Microdiamant), 100\,nm mean width silver cubes (nanoComposix) coated in a <5\,nm polyvinylpyrrolidone (PVP) layer and finally $\sim$10\,$\mu$m long silver wires synthesized in house. The macroscopic silver wires were used as reference markers, during experiment.

\subsection{Experimental characterization}
The sample was mounted on a piezo scan stage, which allowed for identification of single NV-centers by confocal mapping of fluorescence using a 532\,nm linear polarized continuous or pulsed laser with pulse width/period ~50\,ps/400\,ns. The pump light was focused by a 1.4\,NA oil immersion objective, used for excitation and collection of photon, while a halfwave plate controlled the pump polarization. The laser light was filtered from fluorescent photons by dichroic mirrors (SEMROCK) (cut-off 550\,nm) before being detected by a CCD camera (Hamamatsu-Orca-Flash4LT), an EMCCD (Andor - iXon Ultra888) connected to a grating spectrometer (Andor - Shamrock 500i) or two APDs (Picoquant - $\tau$-SPAD) in a Hanbury Brown-Twiss configuration. Photon rate as a function of pump polarization or power, was obtained by the time average of 2\,s time traces accummulated from both APDs.  Decay rate curve and g$^{(2)}$($\tau$) was obtained by histogramming the time interval  from respectively a laser sync pulse or APD detection event, to a detection event on the other APD, using an electronic timing box (Picoquant - PicoHarp 300) in a start-stop configuration. The polarization of emission was probed by the accumulated CCD image count (4s int) for various analyzer orientations. All Experiments were completed within 5 days of spincoating nanocubes.

\subsection{Dark field spectrum}
The dark field spectrum of a single silver cube was obtained in transmission mode, illumination through the glass slide with a 1.2\,NA DF oil condenser lens and collecting light above the sample with a x50 NA 0.75. objective. A 400\,$\mu$m pinhole positioned in image plane in front of a fiber coupled spectrometer, allowed for selection of a single cube. The background signal obtained in the absence of a sample was subtracted from cube- and reference spectrum, using the diffuse scattering from the glass substrate as reference resulting in the cube scattering spectrum. 

\subsection{Finite element modelling}
\begin{figure}[H]
	\centering
		\includegraphics{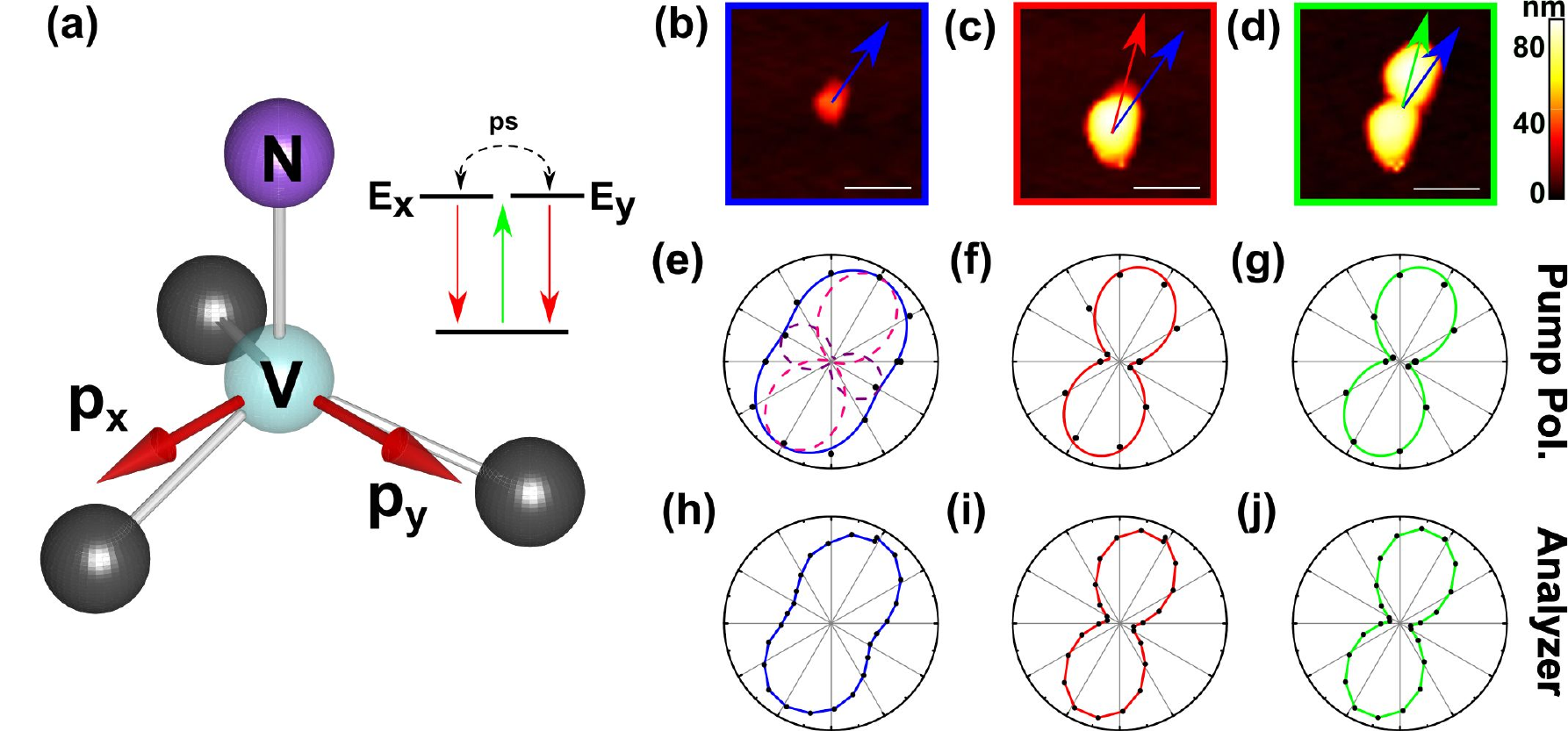}
		\caption{(a) Sketch of the NV-center in diamond. The excitation rate of the NV$^{-}$ charge state scales with the pump field projection on the dipole axes \textbf{p}$_{x}$, \textbf{p}$_{y}$. Phonon promoted population averaging of the doublet excited state E$_x$, E$_y$ (inset) facilitate spontanoues emission, polarized along either dipole axis. (b-d) AFM assembly of photon source color coded to corresponding experimental measurements, for (b) isolated ND (blue), (c) single cube (red) and (d) 2 cube configuration(green) scale bar 200\,nm. Blue arrow represent the orientation of largest inplane dipole moment of the NV-center, while red and green arrow give the electric field orientation of the antenna mode. (e-g) Normalized photon rate vs pump polarization angle, measured (dot) and model fit (solid). Dash curves in (e) indicate potential linear contributions of \textbf{p}$_{x}$ and \textbf{p}$_{y}$ to the model. (h-j) Normalized photon rate vs analyzer angle, measured (dot) and interpolation (solid).}
	\label{fig:Figure2}
\end{figure}

\begin{figure}[H]
	\centering
		\includegraphics{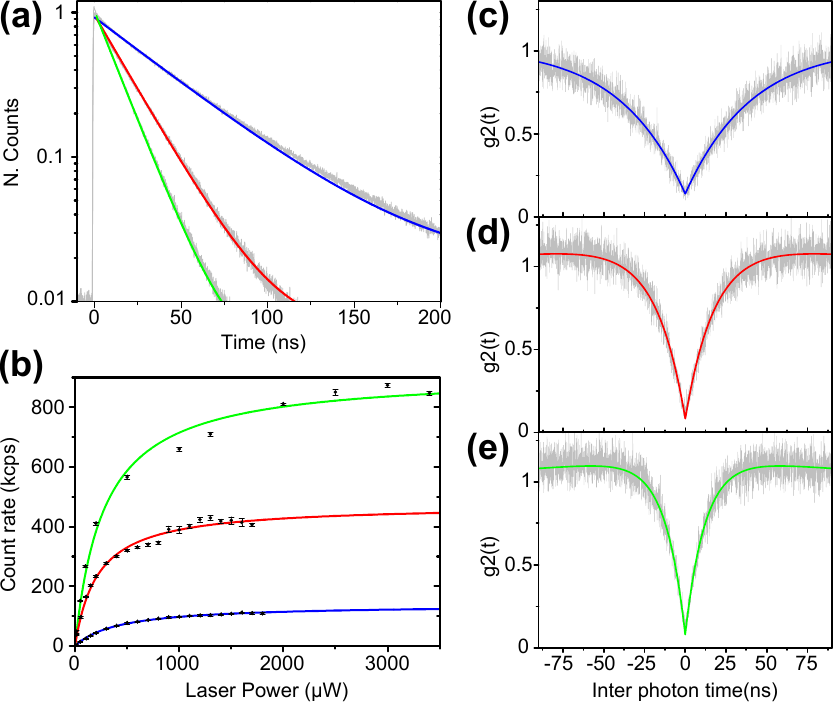}
		\caption{Color coded model fits to experimental data(gray) for isolated ND(blue), single cube(red) and 2 cubes(green). (a) excited state decay curves, fitted to single exponentials. (b) Saturation curves, background corrected for signal from nearby plain surface and (c-e) 2. order correlation measurement (raw data).}
	\label{fig:Figure3}
\end{figure}

\begin{figure}[H]
	\centering
		\includegraphics{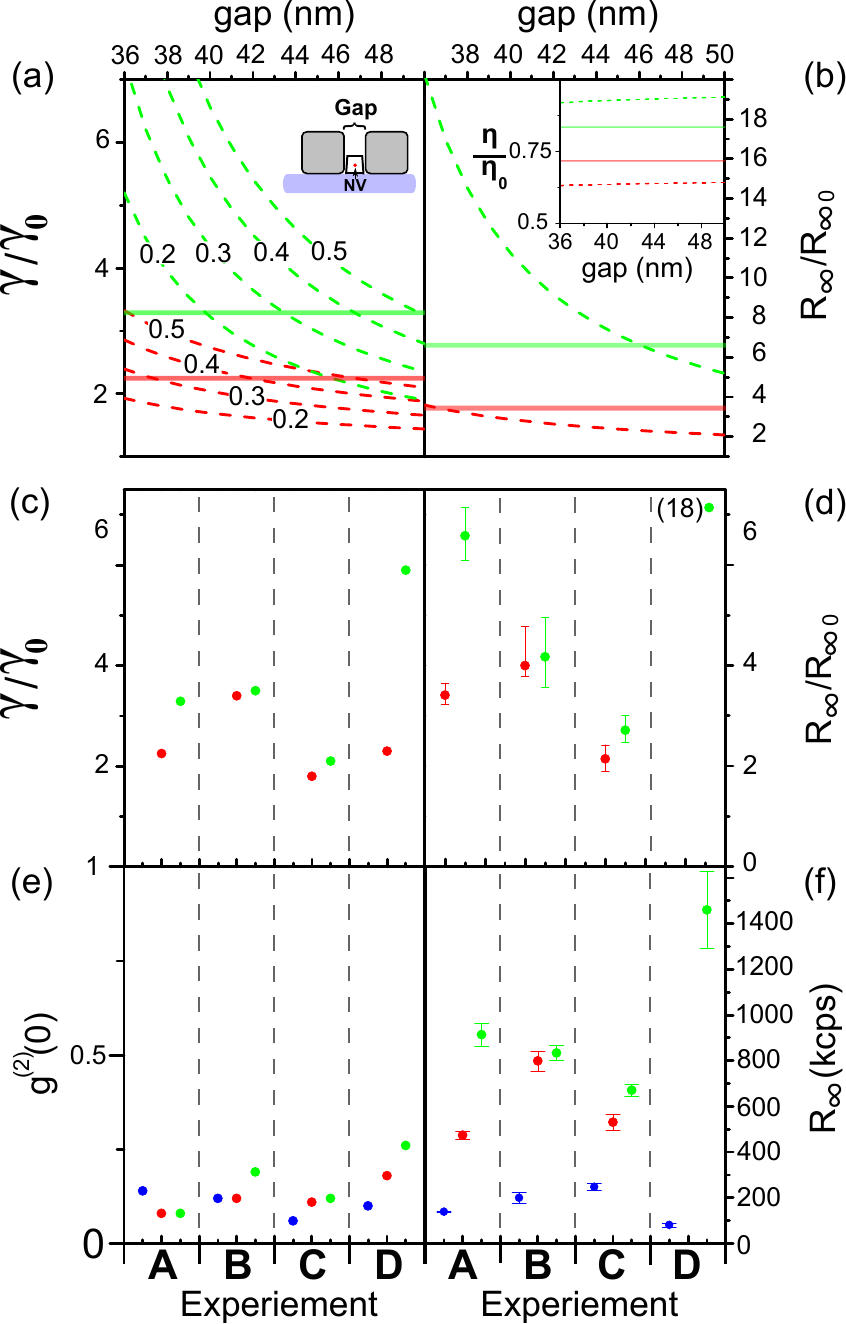}
		\caption{ (a,b) Modelling of experiment A(dashed) and experimental values (solid, horizontal) for a single (red) or two cube configuration (green). (a,inset) Modelled values are given for an NV-center situated in the center of a 35\,nm ND for realistic gap separation of nano cubes. (a) Decay rate enhancement for intrinsic quantum efficiency q$_{e0}$=0.2-0.5 labelled on the curve. (b) Enhancement of detected photon rate at saturation, inset gives the experimentally predicted ratio of collection efficiency for q$_{e0}=0.35$ (solid) together with modelling (dashed). (c-f) Data summary of experiment A-D, for isolated ND (blue), 1 cube (red) and 2 cubes (green). Errorbars indicate the 95\% confidence interval, not shown for (c) as errors $\sim$ 0.01.}
	\label{fig:Figure4}
\end{figure}

3D finite element modelling was conducted in the commercially available Comsol Multiphysics 5.1. The parameter $\Gamma$ is modelled classically using the relation:  $\Gamma=\gamma_r/\gamma_{r0}=P^{NV}/P_0^{NV}$. P$^{NV}$ being the total power dissipated by two orthogonal classical electric dipoles, lying in the dipole plane experimentally determined by the model fit in figure \ref{fig:Figure2}e. wrt. to the antenna axis $\varphi_p^{cube}$.  Simulation performed at emission wavelength 680\,nm, were found to be independent of the dipoles orientation in the plane. Index 0 refer to the reference system of dipoles situated in the center of a 35\,nm tall, 4-sided truncated pyramid shaped, diamond on a semi-infinite glass substrate, bordered from above by an air hemisphere. The modelled domain is bounded by a perfectly match layer (PML). A 8nm corner/side rounding radius were used for the 80\,nm cubes introduced symmetrically around the diamond. Material parameters were based on interpolation of tabulated data\cite{JC,OpticalDiamondProperties,FusedQuartz}. The collection efficiency was modelled as $\eta=P_{obj}^{NV}/P^{NV}$.  P$_{obj}^{NV}$ being the power integrated over a spherical surface in the glass substate, corresponding to the solid collection angle of the experimental 1.4\,NA objective.

\begin{acknowledgement}
The authors gratefully acknowledge the financial support of the European Research Council (Grant 341054 (PLAQNAP)).

\end{acknowledgement}

\begin{suppinfo}

Section S1: Schematic of experimental setup. Section S2: Fluorescence spectra of respectively isolated ND, 1 cube and 2 cube configuration for experiment A. Section S3: Detailed explanation of the model, fitted to photon rate as function of pump polarization angle, for isolated ND and cube coupled configurations. Section S4: Experimental data plots for experiment B-D.

\end{suppinfo}


\providecommand{\latin}[1]{#1}
\makeatletter
\providecommand{\doi}
  {\begingroup\let\do\@makeother\dospecials
  \catcode`\{=1 \catcode`\}=2\doi@aux}
\providecommand{\doi@aux}[1]{\endgroup\texttt{#1}}
\makeatother
\providecommand*\mcitethebibliography{\thebibliography}
\csname @ifundefined\endcsname{endmcitethebibliography}
  {\let\endmcitethebibliography\endthebibliography}{}


\end{document}